\documentclass[preprint2]{aastex}
\usepackage{subfigure}
\usepackage{wrapfig}

\begin{document}

\title{Analyzer of Spectra for Age Determination (ASAD) - Algorithm and Applications}

\author{Randa S. Asa'd}
\affil{American University of Sharjah, Physics Department, P.O.Box 26666, Sharjah, UAE}
\email{raasad@aus.edu}

 \begin{abstract}
 
 $\underline A$nalyzer of $\underline S$pectra for $\underline A$ge $\underline D$etermination (ASAD) is a new package that can easily predict the age and reddening of stellar clusters from their observed optical integrated spectra by comparing them to synthesis model spectra. 
 The ages obtained with ASAD are consistent with ages obtained from previous cluster age methods requiring a more rigorous and time-consuming analysis. This package not only provides fast results, but also allows the user to comprehend the accuracy of these results by providing surface plots and spectral plots for all combinations of observations and models. ASAD is available for download on the Web and can be immediately used on both Mac and Windows.
 
  \end{abstract}  
 

 \section{Introduction}

	Star clusters are the building blocks of galaxies, and the best tools for studying the formation history of these galaxies. When the ages of star clusters are well known, they can reveal the full picture of the different phases in the star-formation history of the host galaxy. 
 There are different methods of determining the ages of stellar clusters. The one that is considered to be the most accurate is the method of the color-magnitude diagram (CMD), in which the constituent stars of the cluster are each individually measured. For far away galaxies, however, one cannot observe the individual stars, that's why providing methods able to determine ages from integrated properties of clusters is important. In \citet{Asad12} a weak correlation between the ages obtained from integrated photometry and the CMD ages was shown. In \citet{Asad13} we showed that the integrated spectra can better predict the age of stellar clusters. 
  Many studies have used integrated spectra of stellar clusters to obtain ages \citep {Santos95, Piatti02, Piatti05, Santos06, Palma08, Talavera10, Beasley02, Wolf07, Koleva08, Fernandes10},  but few user-friendly tools are available to easily obtain ages of clusters instantaneously from their integrated spectra (like Starlight (http://astro.ufsc.br/starlight/), EZ\_age (http://astro.berkeley.edu/~graves/ez\_ages.html), FISA \citep {Benitez-Llambay12} ).  This was the main motivation of this work. I am presenting a new tool that predicts the age and extinction of star clusters from their optical integrated spectra. This tool can be used as an $\underline A$nalyzer of $\underline S$pectra for $\underline A$ge $\underline D$etermination, and hence the name ASAD. ASAD is available for download on the Web and can be immediately used on both Mac and Windows\footnote{http://randaasad.wordpress.com/asad-package/}.
  The algorithm and tests of ASAD are presented in section 2, analysis and applications of what ASAD can do are discussed in section 3. The applications of ASAD are extended beyond the original sample of clusters in section 4. The summary is given in section 5.
  
  \section{Algorithm and Tests}

In order to make ASAD easy to use, the only required input is the integrated spectrum of the cluster. The user can choose the best parameters needed for each study. The basic idea of ASAD is to compare the observed integrated spectra of star clusters with the computed spectral models of \citet {Delgado05} to predict the best age/reddening combination, as introduced in \citet{Asad13}.

 The \citet {Delgado05} models were chosen because of their high spectral resolution (0.3\AA) and the large sample of ages available. This model assumes \citet {Salpeter55} initial mass function (IMF) with cut-off masses of 0.1 - 120 solar masses. The library used in the calculations is the Padova asymptotic giant branch (AGB) stellar model for the metallicities: Z= 0.004, Z= 0.008 and Z= 0.019. More models will be available in future versions of ASAD.
For the sample of the stellar clusters in this work, all from the Large Magellanic Cloud (LMC), Z= 0.008 was used. LMC clusters are known to have [Fe/H] values from -2.2 to 0.0 dex, or between -1.0 to 0.0 dex, for the age range we are using. Our analysis is done on spectra taken in the optical range. \citet {Bica86a} and \citet {Benitez-Llambay12} have shown that metallicity does not play a significant role in the optical range when applying spectral aging methods. This is because optical lines are relatively insensitive to this level of change in metallicity. The stellar clusters sample in this work has ages $<$ 2 Gyr. \citet {cole2005} show that for ages $<$ 5 Gyr most LMC clusters have a [Fe/H] $=$ -0.4.

ASAD first smoothes the model, if needed. The user can pick the wavelength bin preferred as multiples of 0.3\AA. ASAD performs the smoothing by taking the average of the first (2S$-$1) rows, where S is the factor given in Equation 1, then taking the average of the next (2S$-$1) rows starting with the row number (S+1) and so on. Note that when applying this smoothing, the first wavelength in the new created file model is equal to the wavelength in the S$^{th}$ row of the original model file.  

\begin{equation}  S =\frac{\mathrm{wanted \: bin \: size \: in \: \AA}}{0.3}.
\end{equation}  

ASAD reads the observed integrated spectrum from ASCII files. The spectra should be in bins of 0.3\AA \, or multiples of 0.3\AA. The user then specifies the initial and final wavelength as well as the bin size of the wavelength range to be used in the analysis. ASAD smoothes the observation according to the binning size desired. It keeps the original value of the starting wavelength and performs the smoothing as described above. 
ASAD applies the \citet {Cardelli89} extinction law for the optical/NIR range on the observed spectra (with R = 3.1). The user can choose the range of the extinction to be applied in steps of 0.01. The user can also choose the wavelength point where the flux of both the model and observation spectra are normalized. 

All the spectra of the cluster are de-reddened over the range specified, then compared with the models. The model consists of computed integrated spectra starting with log (age/year) 6.60 up to log (age/year) 10.25 in steps of log (age/year) of 0.05. The method used for determining the best match is the $\chi^{2}$ minimization method:

\begin{equation}  \chi^{2} = \sum_{\lambda=\lambda_{min(\AA)}}^{\lambda_{max(\AA)}} (\frac{(OF)_{\lambda} - (MF)_{\lambda}}{(OF)_{\lambda_{arbitrary(\AA)}}})^{2}.
\end{equation}

where OF is the observed flux and MF is the model flux. To test the outputs of ASAD, the package was used to reproduce ages of all the stellar clusters presented in \citet{Asad13}.

\section{Analysis}

\subsection{Plots of Observed Spectra versus Model Spectra}

\begin{figure}
\includegraphics[angle=0,scale=0.4]{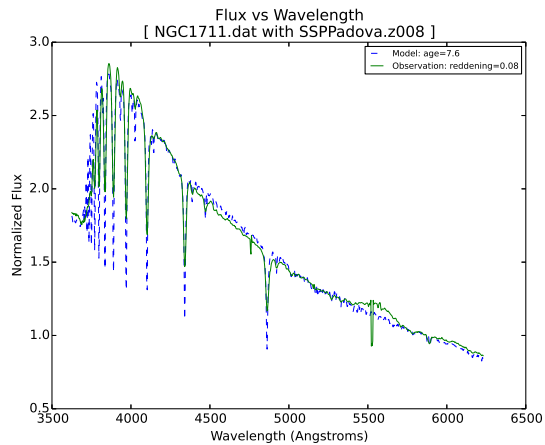}
\caption{ASAD can produce plots of combinations of observations versus models on top of each other for direct comparison.}
 \label{NGC1711a}
\end{figure}

\begin{figure}
\includegraphics[angle=0,scale=0.4]{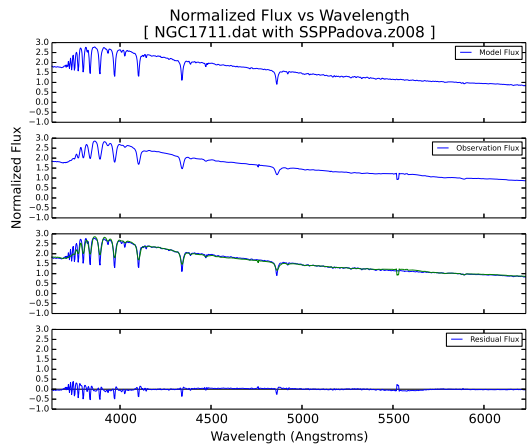}
\caption{ASAD can produce plots showing the residuals of (observation spectrum - models spectrum).}
 \label{NGC1711b}
\end{figure}

After calculating the best age/reddening combination obtained from the $\chi^{2}$ minimization method, ASAD can plot the spectra of the best match so that the user can check the accuracy of the result. An application showing the importance of this plot is shown later in Section 4.1. 
The user can also choose other combinations of reddening and ages to be plotted and compare them with the predicted output. 
There are two formats available for the user to use when plotting the spectra. They can be plotted on top of each other for direct comparison as shown in Figure \ref{NGC1711a} or they can be plotted separately in addition to their residuals, where the residual is found by subtracting the model spectrum form the observation spectrum bin by bin as shown in Figure \ref{NGC1711b}.

\subsection{Reddening Limit}

In \citet{Asad12} a sample of 84 resolved clusters was analyzed. In the literature, none of these clusters had reddening values greater than 0.35, however, we showed that using an integrated photometry method to solve for age and reddening simultaneously, inflated reddening solutions were found. The upper limit for reddening used in that work was E(B-V) = 0.8 (see the paper for more details). In \citet{Asad13} where we used an aging method of integrated spectra for a sample of 27 clusters, we set the reddening limit to E(B-V) = 0.5. The predicted reddening values  were less than E(B-V) = 0.3. 
In this work I used ASAD to test the predicted results of the integrated spectra method when the reddening limit is increased to E(B-V) = 0.8. The results were not affected by this increased reddening limit. The limit was then stretched to E(B-V) = 1.5 but the results were still the same. No inflated reddening solutions are found like with broad-band methods. 

\subsection{Surface Plots}

\begin{figure}
\includegraphics[angle=0,scale=0.6]{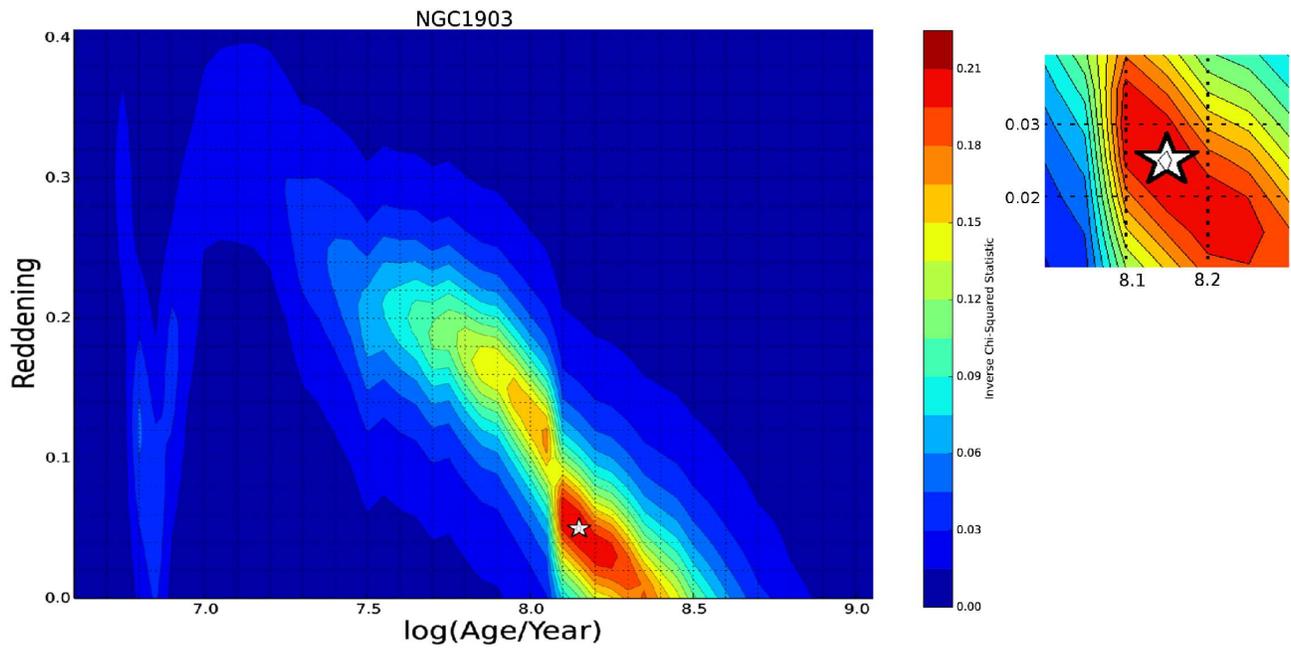}
\caption{The surface plot of the age-reddening combinations for the cluster NGC1903. An example of the ideal case with a single-valued and precise solution. Note that when zooming in the Figure, the region of the accepted solutions is all included within the star.}
 \label{NGC1903surface_a_c}
\end{figure}

\begin{figure}
\includegraphics[angle=0,scale=0.4]{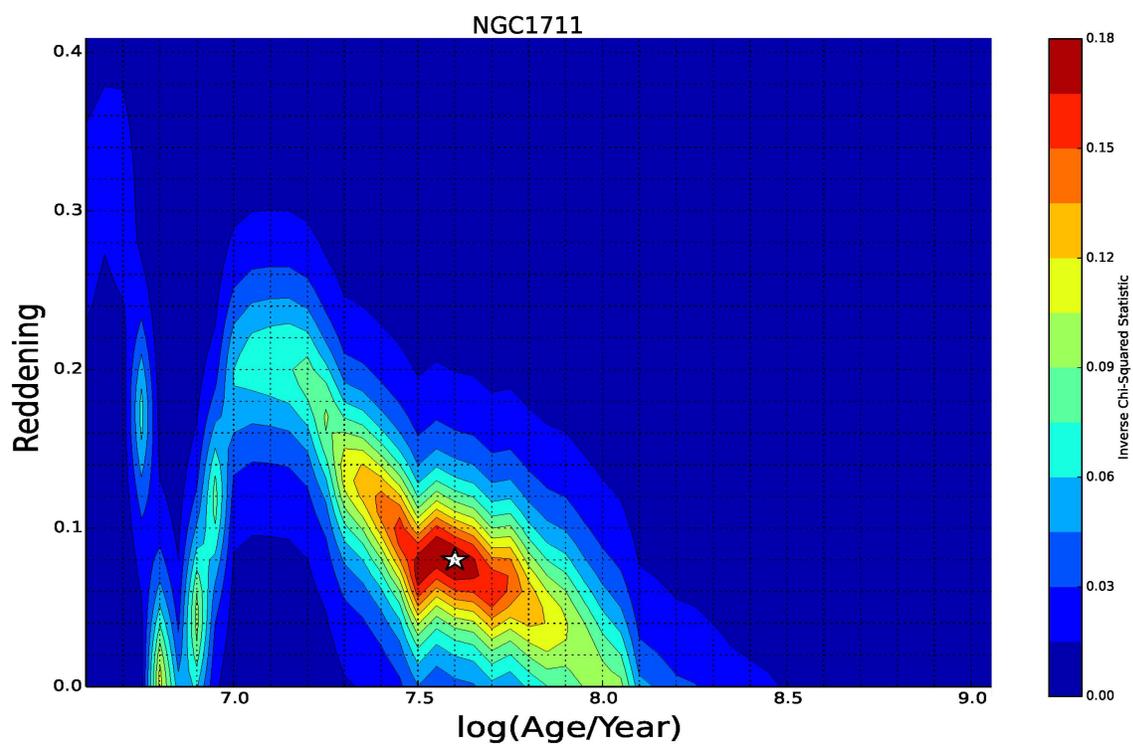}
\caption{The surface plot of the age-reddening combinations for the cluster NGC1711. The solution is unique but not very precise.}
 \label{NGC1711surface_c}
\end{figure}

\begin{figure}
\includegraphics[angle=0,scale=0.4]{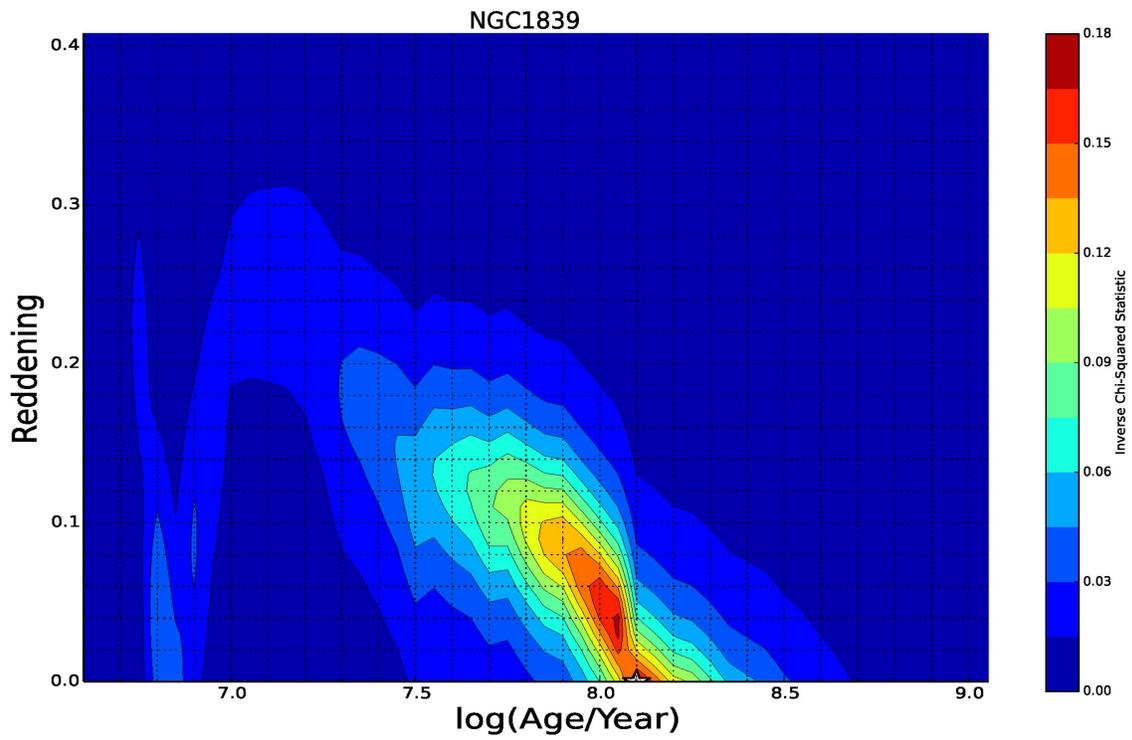}
\caption{The surface plot of the age-reddening combinations for the cluster NGC1839. There could be two possible solutions for the assigned age-reddening values.}
 \label{NGC1839surface_c}
\end{figure}

To better understand the results of the $\chi^{2}$ minimization method, ASAD can create plots showing the inverse of the $\chi^{2}$ values over a 2D surface plot of the full age and reddening range explored by the algorithm. These surface plots provide information about the uniqueness of the solution (the age/reddening values assigned for the cluster) as well as the precision of the result (a widely spread surface of possible solutions around the assigned solution means a less precise result). 
ASAD was used to create the surface plots for our sample of 27 clusters described in \citet{Asad13}. Three of them are shown in Figures \ref{NGC1903surface_a_c}, \ref{NGC1711surface_c} and \ref{NGC1839surface_c}. Figure \ref{NGC1903surface_a_c} (NGC1903) is an example of the ideal case with a single-valued, precise solution. Note that when zooming in, the region of the best solutions is all included within the star. Figure \ref{NGC1711surface_c} (NGC1711) shows a case in which the solution is unique but not very precise. Figure \ref{NGC1839surface_c} (NGC1839) shows a case where there are two minima, meaning that there could be two possible solutions for the assigned age/reddening values. Luckily, this case happens only twice in our sample of 27 clusters, but it is an alarm for the need to study the 2D surface plot of each cluster individually in order to check the reliability of the obtained results.

\subsection{Metallicity}

As mentioned in Section 2, \citet {Bica86a} and \citet {Benitez-Llambay12} have shown that metallicity does not play a significant role in the optical range when applying spectral aging methods.

\begin{figure}
\includegraphics[angle=0,scale=0.7]{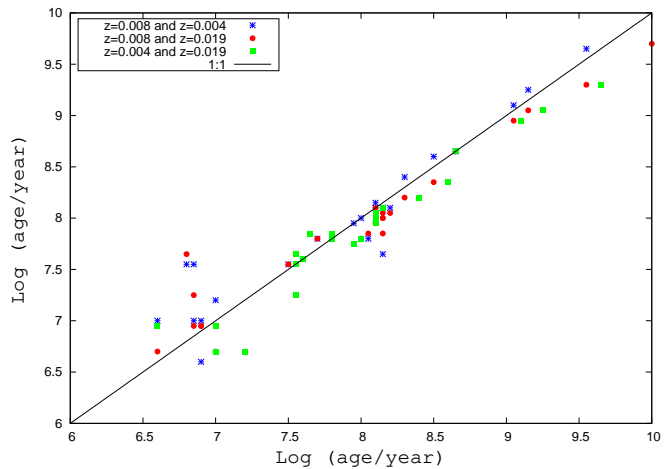}
\caption{The age prediction using different combinations of metallicity. The blue stars show log (age/year) obtained using metallicity Z= 0.004 versus log (age/year) obtained using metallicity Z= 0.008. The red circles show log (age/year) obtained using metallicity Z= 0.019 versus log (age/year) obtained using metallicity Z= 0.008. The green squares show log (age/year) obtained using metallicity Z= 0.019 versus log (age/year) obtained using metallicity Z= 0.004.}
 \label{Metal}
\end{figure}

To test this, I ran ASAD on my sample using the model with the three metallicities available Z= 0.004, Z= 0.008 and Z= 0.019 to obtain the ages. Figure \ref{Metal} shows the results using the different combinations of metallicity. The blue stars show log (age/year) obtained using metallicity Z= 0.004 versus log (age/year) obtained using metallicity Z= 0.008. The red circles show log (age/year) obtained using metallicity Z= 0.019 versus log (age/year) obtained using metallicity Z= 0.008. The green squares show log (age/year) obtained using metallicity Z= 0.019 versus log (age/year) obtained using metallicity Z= 0.004. The correlation coefficient for each combination is 0.96, 0.97, 0.98 respectively. Except few outliers, the results are close to the 1:1 line, showing that different metallicities do not have a significant effect on the age prediction when using the full optical spectrum matching method.

\clearpage

\clearpage
\clearpage

\clearpage

\clearpage

\section{Applications Beyond our Sample}

\subsection{Spectral Templates}

Spectral templates \citep {Bica90, Santos95, Piatti02} have been widely used to obtain the age and reddening of stellar clusters. They were created from real observations of clusters with known age and extinction properties. 
ASAD was used to investigate the relation between the ages assigned to these templates and the ages predicted by the SSP models, using the models of \citep {Delgado05}. 
Note that the ages predicted by some of those templates are not single-valued but rather a range of ages. For the analysis I took the average age value within that range. Young and intermediate templates were tested. 
The templates are given in wavelength steps of 2\AA, and the model in steps of 0.3\AA. To match the templates with the model they were both smoothed to steps of 6\AA. 
Those templates cover different wavelength ranges. To perform a consistent comparison, a common wavelength range of 3656-6230\AA \, was used. The templates are already corrected for reddening, thus the only variable in our analysis for this section is the age. It is worth mentioning that the reddening correction of the templates is done using \citet {Seaton79} extinction law rather than the \citet {Cardelli89} extinction law used in our analysis. 

\begin{figure}
\includegraphics[angle=0,scale=0.5]{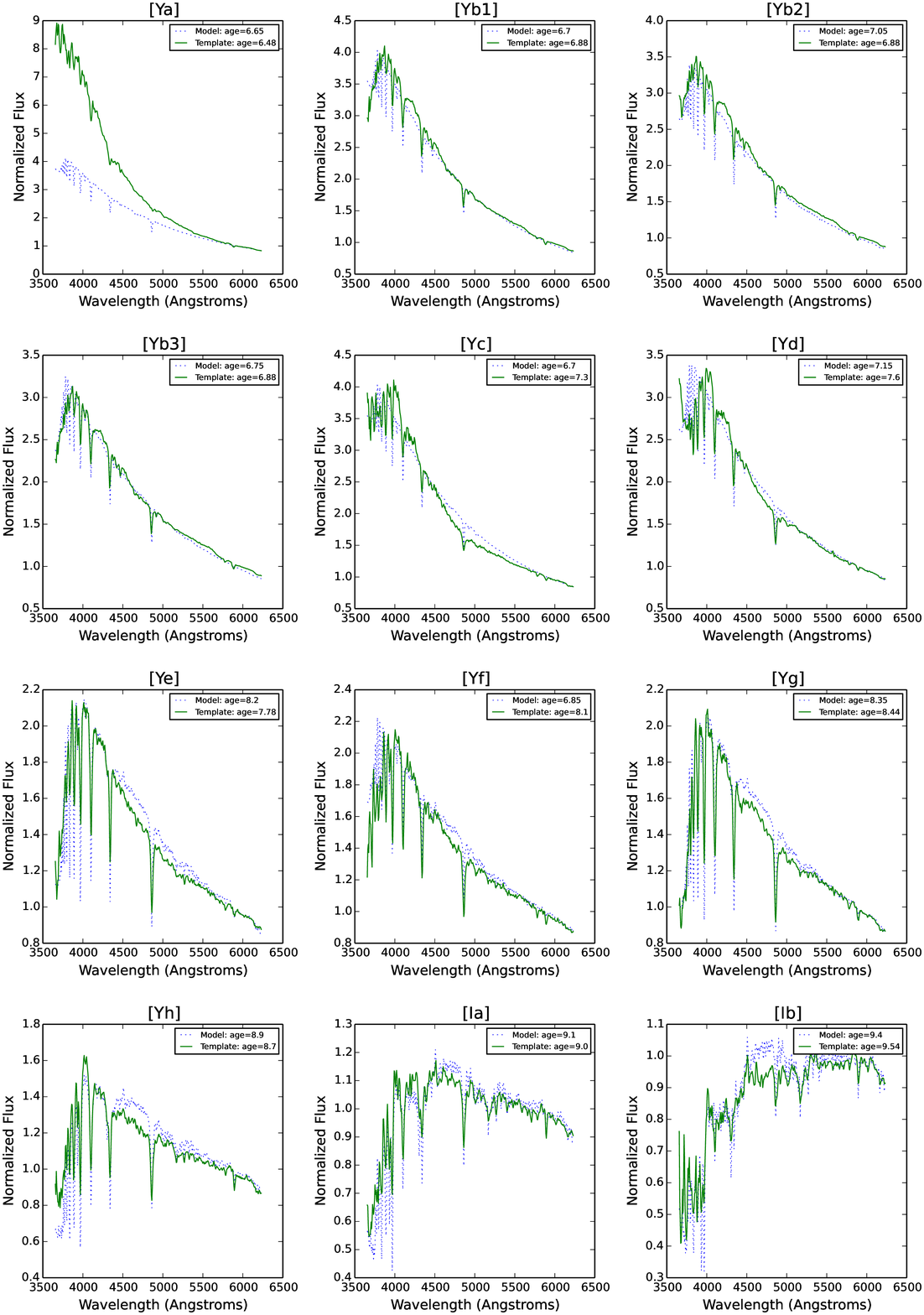}
\caption{The combinations of the best match between templates and models for all templates.}
\label{TemplatesCombinations_a}
\end{figure}

The combinations of the best match between templates and models is shown in Figure \ref{TemplatesCombinations_a}. The following points are noticed:
1. (Ya) template has a flux greater than the youngest synthetic spectral model can predict. The age assigned for this template is the age of the synthetic spectrum with highest flux among the model spectra, that is log (age/year) 6.65.
2. There is a reoccurring continuum gap in almost all the matches of Figure \ref{TemplatesCombinations_a} in the range 4500-4800\AA \,, near the H$_{\beta}$ line. 
3. Visual inspection shows that the (Yf) template belongs to a younger cluster than is assigned. Balmer lines are even weaker than (Ye) but stronger than (Yd). Visually, it appears to have a log (age/year) of perhaps 7.4.

\begin{figure}
\includegraphics[angle=0,scale=0.7]{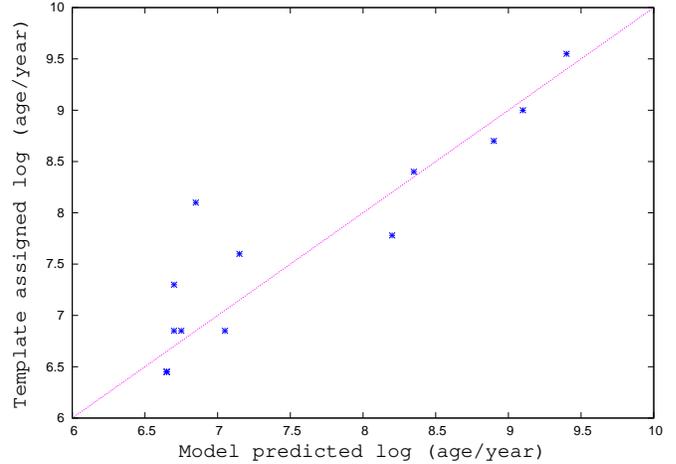}
\caption{The ages of the templates versus the ages predicted for those templates from the \citet {Delgado05} model. The correlation coefficient is 0.91}
\label{TemplatesModelNoRed_eps}
\end{figure}

\begin{figure}
\includegraphics[angle=0,scale=0.7]{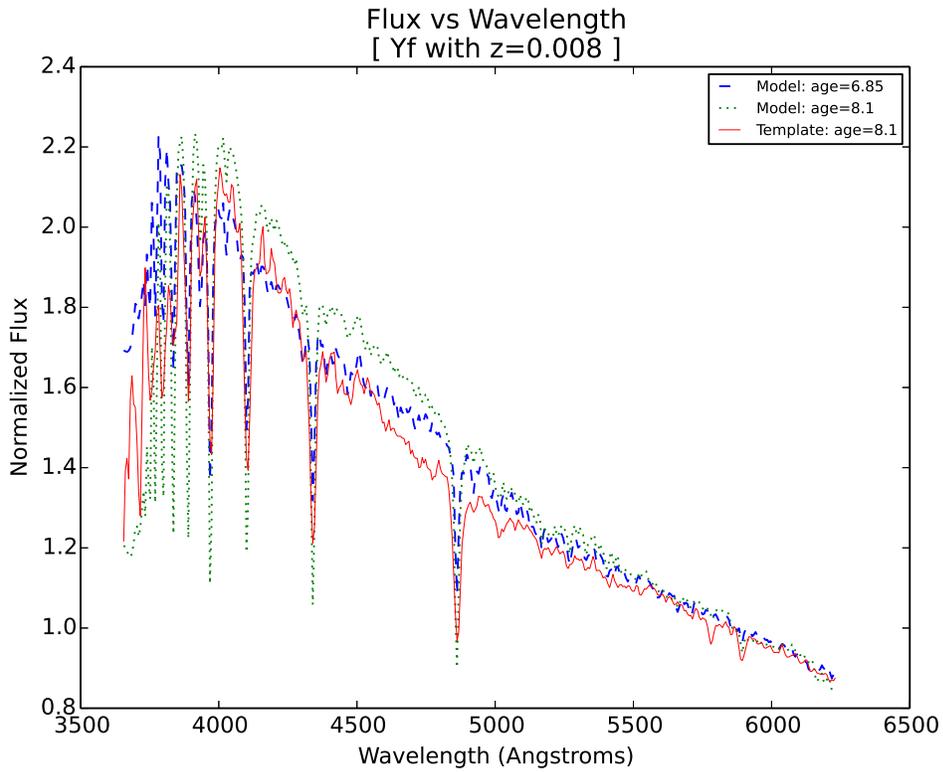}
\caption{Yf template with the model synthetic spectra for both log (age/year) 6.85 and 8.1. Note that in the range 3860-6230\AA \,the model spectrum for log (age/year) 8.1 is a better fit, but what makes the spectrum of log (age/year) 6.85 a better fit overall is the region 3656-3860\AA \, where it is closer fit to the template spectrum.}
\label{Yf}
\end{figure}

The results are shown in Table \ref{Templates} and Figure \ref{TemplatesModelNoRed_eps}. Note that the numeric character in Ya and Yb templates in Table \ref{Templates} is to differentiate cluster spectra corrected by different reddening values. The correlation coefficient is 0.91. (Yf) is an outlier.  The template log (age/year) is 8.1, but our predicted log (age/year) is 6.85. Figure \ref{Yf} shows this template, with the model synthetic spectra for both log (age/year) 6.85 and 8.1. Note that in the range 3860-6230\AA \, the model spectrum for log age 8.1 is a better fit, but what makes the spectrum of log age 6.85 a better fit overall is the region 3656-3860\AA \, where it is closer fit to the template spectrum. 
As a test, the wavelength range towards the red was stretched up to 6980\AA, and we got the same results. This implies that the blue side of the spectrum is more age-sensitive than the tail of the red side of the spectrum. 

\clearpage
\clearpage

\begin{deluxetable}{|c|c|c|c|}
\tabletypesize{\scriptsize}
\tablecaption{Ages of Templates  \label{Templates}}
\tablewidth{0pt}
\tablehead{
\colhead{Name} &  \colhead{Range of Template log (Age/year)} &  \colhead{Adopted Template log (Age/year)} &  \colhead{Predicted log (Age/year) using ASAD}
}
\startdata

Ya1 & 6.30 - 6.60   & 6.45 & 6.65\\
Ya2 &  6.30 - 6.60  & 6.45 & 6.65\\
Ya3 &  6.30 - 6.60   &  6.45 & 6.65\\
Yb1 &  6.70 - 70 & 6.85 & 6.70\\
Yb2 &  6.70 - 7.00  & 6.85 & 7.05\\
Yb3 &  6.70 - 7.00 &  6.85 & 6.75\\
Yc &  7.30  & 7.30 & 6.70\\
Yd &  7.60  & 7.60 & 7.15\\
Ye &  7.65 - 7.90  & 7.78 & 8.20\\
Yf &  8.00 - 8.20  & 8.10 & 6.85\\
Yg &  8.30 - 8.50  & 8.40 & 8.35\\
Yh &  8.70 & 8.70 & 8.90\\
Ia & 9.00  &  9.00 & 9.10\\
Ib &   9.50 - 9.60  & 9.55 & 9.40\\

\enddata
\end{deluxetable}

\subsection{Literature Clusters}

\begin{figure}
\includegraphics[angle=0,scale=0.7]{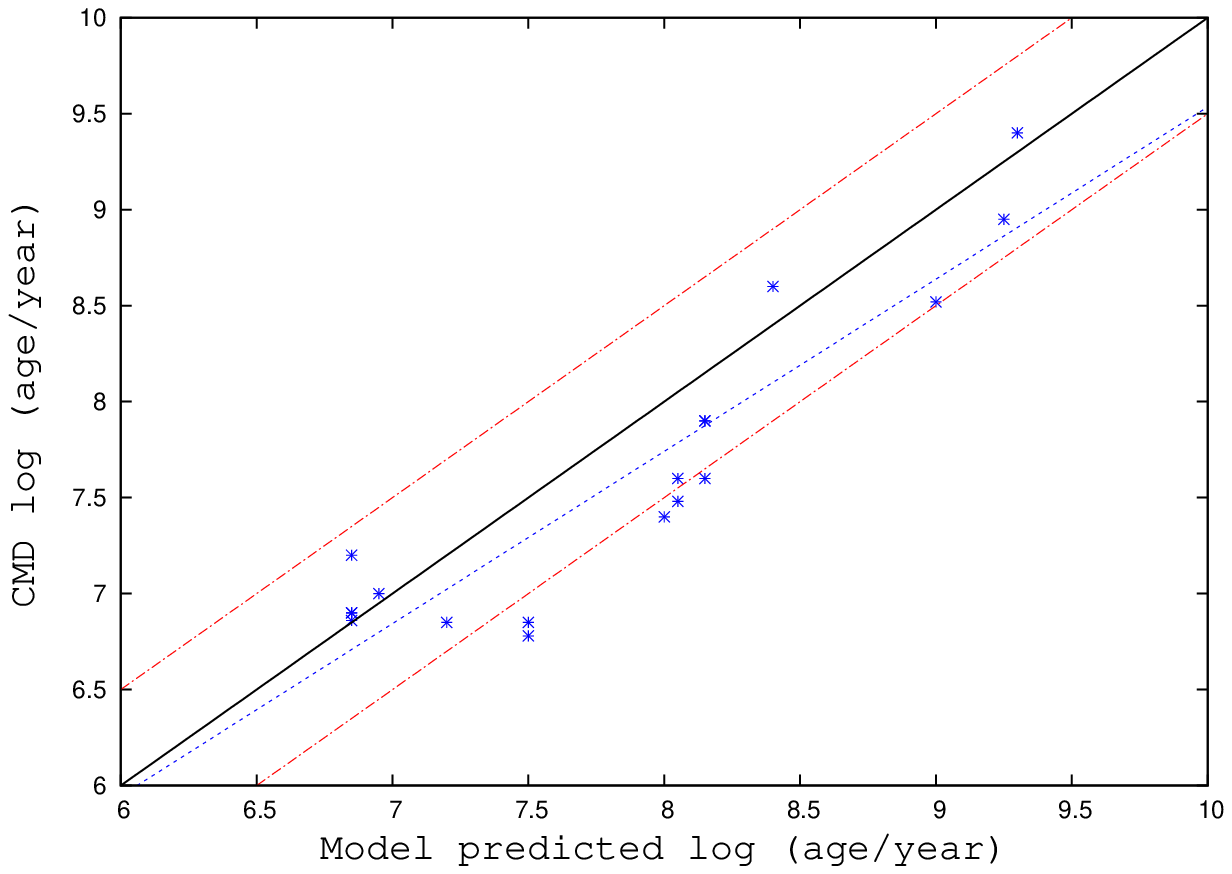}
\caption{The CMD ages of the literature clusters versus the ages predicted for those clusters with ASAD. Two red dotted lines represent the range of  +/- 0.5 in log (age/year). The blue line is the fit line. The  correlation coefficient is 0.92.}
\label{LitClusters}
\end{figure}

 ASAD was used to predict the ages of LMC stellar clusters with publicly available spectra\footnote{http://vizier.cfa.harvard.edu/viz-bin/VizieR-3?-source=III/219/cluster}. Despite the large number of integrated spectra, a subsample of only 18 spectra was used because of two factors. First, to be able to test the accuracy of the results obtained with ASAD, I only picked the spectra of clusters that have CMD ages in the literature. Second, to make the analysis consistent, I chose a common wavelength range available for as many clusters as possible. The range of 3692-5852\AA \, was used and the flux of both the model and the data was normalized at 5852\AA. As done for the templates, the step size of 6\AA \, was used to match the data to the models because the data is given in steps of 2\AA.   
The results are shown in Table \ref{LiteratureClusters} and Figure \ref{LitClusters}. The correlation coefficient between the ages obtained with ASAD for these clusters, and their CMD ages is 0.92. Four of the five outliers were part of the sample of clusters in \citet{Asad13}. The results obtained for them using ASAD were better. This means that the accuracy of the aging is could be limited by the quality of the observed spectrum rather than the method used. 

\begin{deluxetable}{|c|c|c|}
\tabletypesize{\scriptsize}
\tablecaption{Ages of Literature Clusters  \label{LiteratureClusters}}
\tablewidth{0pt}
\tablehead{
\colhead{Name} &  \colhead{CMD log (Age/year)} &  \colhead{Predicted log (Age/year) using ASAD}
}
\startdata

NGC1711 & 7.40 & 8.00\\
NGC1783 & 8.95 & 9.25\\
NGC1831 & 8.60 & 8.40\\
NGC1847 & 7.20 & 6.85\\
NGC1850 & 7.60 & 8.05\\
NGC1854 & 7.48 & 8.05\\
NGC1856 & 7.90 & 8.15\\
NGC1866 & 7.90 & 8.15\\
NGC1868 & 8.52 & 9.00\\
NGC1978 & 9.40 & 9.30\\
NGC1983 & 6.90 & 6.85\\
NGC1984 & 6.85 & 7.50\\
NGC1994 & 6.86 & 6.85\\
NGC2004 & 6.90 & 6.85\\
NGC2011 & 6.78 & 7.50\\
NGC2100 & 7.00 & 6.95\\
NGC2102 & 6.85 & 7.20\\
NGC2157 & 7.60 & 8.15\\

\enddata
\end{deluxetable}

\section{Summary}

The ASAD spectral analysis package has been introduced and demonstrated on previously published integrated spectra of stellar clusters. I presented results showing an improved method for finding the ages of stellar clusters by means of integrated spectra. We summarize the findings we obtained using ASAD in the following points:

1. The method of obtaining the ages of stellar clusters by fitting their observed spectra to the theoretical SSP model spectra is a reliable method.\\
2. Examining the surface plots of the age/reddening space when performing the $\chi^{2}$ minimization calculation is essential for understanding the uniqueness and range of the solution obtained. \\
3. Examining the best spectral match between the models and observations is important in understanding the reliability of the obtained results. \\
4. Extending the reddening limit beyond E(B-V) =0.5 for the LMC clusters will not affect the results obtained.\\ 
5. There is a reasonable match between the ages of the templates and the ages predicted by the \citet {Delgado05} for these Templates (using ASAD) except a few. More investigation is needed to understand the limitation of using templates and synthetic models. \\
6. The blue side of the optical spectrum is more sensitive than the red side for age determination of stellar clusters.\\
7. An aging limitation might be caused by the quality of the observed spectrum rather than the method itself. \\

\section{Acknowledgement}

I would like to thank Dr. Margaret Hanson for carefully reviewing the manuscript of this paper and providing extensive, critical comments. 
I would like to acknowledge the significant contribution of the student Sami Abdin who developed the ASAD program code. I also thank A. M. Asad who worked on the first version of ASAD. The students Adnan Shahpurwala and Hala Sarhan contributed to parts of this work, too. 
This material is based upon work supported in part by the Seed Grant and FRG3 Grant P.I., R.\ Asa'd from American University of Sharjah. This work was also supported in part by the National Science Foundation under Grant No.\ AST-1009550.

\clearpage
\clearpage

\bibliographystyle{apj}     
\bibliography{LatestDec25}

\end{document}